\documentclass[12pt]{iopart}
\usepackage{graphicx}   % Include figure files
\usepackage{bm}         % bold math
\usepackage{textcomp}
\usepackage{psfrag}
\usepackage{iopams}
\usepackage[mathscr]{eucal}

\newcommand{\beqs}{\begin{eqnarray*}}
\newcommand{\eeqs}{\end{eqnarray*}}
\newcommand{\beq}{\begin{eqnarray}}
\newcommand{\eeq}{\end{eqnarray}}
\newcommand{\bit}{\begin{itemize}}
\newcommand{\eit}{\end{itemize}}

\newcommand{\np}{n^+}
\newcommand{\nm}{n^-}
\newcommand{\p}{\partial}

\newcommand{\dm}{\lambda_m}

\newcommand{\kp}{\mathbf{k}_\perp}
\newcommand{\kpm}{k_\perp}
\newcommand{\rp}{\mathbf{r}_\perp}

\newcommand{\fp}{\mathbf{f}_\perp}
\newcommand{\epsm}{\epsilon_m}

\begin{document}

\title{A Poisson-Boltzmann approach for a lipid membrane in an electric field}
\author{Falko Ziebert and David Lacoste}
\address{Laboratoire de Physico-Chimie Th\'eorique - UMR CNRS Gulliver 7083,
ESPCI, 10 rue Vauquelin, F-75231 Paris, France}
\ead{david@turner.pct.espci.fr}
\begin{abstract}
The behavior of a non-conductive quasi-planar lipid membrane in an electrolyte and 
in a static (DC) electric field is investigated theoretically
in the nonlinear (Poisson-Boltzmann) regime. Electrostatic effects
due to charges in the membrane lipids and in the double layers lead 
to corrections to the membrane elastic moduli which are analyzed here.
We show that, especially in the low salt limit,
i) the electrostatic contribution  to the membrane's surface tension
due to the Debye layers crosses over
from a quadratic behavior in the externally applied voltage to a linear voltage regime.
ii) the contribution  to the membrane's bending modulus due to the Debye layers saturates for
high voltages.
Nevertheless, the membrane undulation instability due to an effectively negative surface tension
as predicted by linear Debye-H\"uckel theory
is shown to persist in the nonlinear, high voltage regime.
\end{abstract}

\pacs{87.16.-b, 82.39.Wj, 05.70.Np}
\submitto{\NJP}
\maketitle

%%%%%%%%%%%%%%%%%%%%%%%%%%%%%%%%%%%%%%%%%%%%%%%%%%%%%%%%
\section{\label{Intro}Introduction}
%%%%%%%%%%%%%%%%%%%%%%%%%%%%%%%%%%%%%%%%%%%%%%%%%%%%%%%%

Bilayers formed from lipid molecules are an essential component
of the membranes of biological cells. The mechanical properties
of membranes at equilibrium are characterized by
two elastic moduli, the surface tension and the bending modulus \cite{seifert_mb_review:1997}.
These moduli typically depend on electrostatic properties,
and their modifications in case of charged membranes/surfaces in an electrolyte
have been examined theoretically in the period 1980-90's as reviewed e.g.~in Ref.~\cite{andelman}:
they have been first derived by Winterhalter and Helfrich \cite{Helf88} within linearized Debye-H\"uckel (DH) approximation,
then by Lekkerkerker \cite{Lekk89} in the nonlinear Poisson-Boltzmann (PB) regime
for charged monolayers and by Ninham et al.~\cite{Ninham89} for charged symmetric bilayers.
Later on Helfrich et al.~revisited the question of the electrostatic corrections
to the bending modulus of charged symmetric bilayers \cite{WinterHelf92}.

Nowadays, the study of deformations of membranes or vesicles in electric fields is
an active field of research
linked to many biotechnological applications.
For instance, the application of electric fields is used to produce artificial vesicles
from lipid films (electroformation),
or to create pores in vesicles (electroporation), which is an important route for drug delivery.
Both processes are widely used experimentally although
they are not well understood theoretically. The
effects of electric fields on giant unilamellar vesicles
have been reviewed recently in Ref.~\cite{dimova:09}.
This system shows a rich panel of possible behaviors and morphological transitions
depending on experimental conditions -- electric field frequency, conductivities of the
medium and of the membrane, salt concentration, etc.
These observations are supported to a large extend by theoretical modeling \cite{petia:2009}.

Originally it was Helfrich et al.~who pointed out that the deformation of lipid vesicles in electric fields
can serve as a means to determine
the electrostatic corrections to the membrane elastic moduli \cite{WinterHelf91}.
Besides this observation on the macroscopic scale
other techniques can provide valuable insights into the moduli corrections such as AFM,
impedance spectroscopy \cite{sackman:2002},
neutron reflectivity \cite{charitat_EPJE} and X-ray scattering \cite{charitat_2009}.
Recently \cite{charitat_2009}, X-ray scattering experiments have been carried out
on a system of two superposed lipid membranes in an AC electric field.
In trying to analyze this data, we noted that these experiments have been carried out
at relatively high voltages, in a regime where the linearized DH
approach may not be applicable.
In order to describe such a situation theoretically, we extend previous work
\cite{Lacoste_EPL,Lacoste_EPJE,ZBL} based on the DH approach
to the nonlinear PB regime, which is more suitable for realistic situations
in which the induced surface charges on the membrane are large.

In this paper, we present a simple approach to calculate
electrostatic corrections in the elastic moduli of a quasi-planar lipid 
membrane. The membrane
is assumed to be non-conductive to ions, 
non-permeable to water and electrically neutral, it is subjected to a normal
DC electric field and embedded in an electrolyte described by the Poisson-Boltzmann equation.
In this situation, the electric field leads to an accumulation of charges
on both sides of the membrane, which affect the mechanical properties of the membrane.
The electrostatic corrections to the elastic moduli can be used to 
analyze the instability of a lipid membrane in an applied DC electric field.
In contrast to previous work, based on a free energy approach
\cite{Lekk89,WinterHelf92,lomholt_elect}, our method is
based on a calculation of the stress (or force) balance at the membrane surface
using electrokinetic equations \cite{armand}. This method is related 
to the work of Kumaran \cite{kumaran} who used a similar approach
 in the context of equilibrium charged membranes.
Two points are worth mentioning:
first, our approach is able to describe the capacitive effects
of the membrane and of the Debye layers while keeping the simplicity of the zero thickness
approximation on which most of the literature on lipid membranes is based.
Second, our approach can include non-equilibrium effects
which can not be described within the free energy approach.
For instance, in Refs.~\cite{Lacoste_EPL,Lacoste_EPJE,ZBL} we investigated
the effects of ionic currents flowing through the membrane, 
which in turn affect the fluid flow near the membrane.
Other types of non-equilibrium effects that could be included
in that framework are those arising from the stochasticity of ion channels.

%%%%%%%%%%%%%%%%%%%%%%%%%%%%%%%%%%%%%%%%%%%%%%%%%%%%%%%%
\section{\label{Model}Model equations}
%%%%%%%%%%%%%%%%%%%%%%%%%%%%%%%%%%%%%%%%%%%%%%%%%%%%%%%%

We consider a steady (DC) voltage $V$ between two electrodes at a fixed distance $z=\pm L/2$,
applied to an initially flat membrane located at $z=0$.
The membrane is embedded in an electrolyte of monovalent ions with densities $n^+$
and $n^-$.
The membrane is treated as non-conductive for both ion species and is (effectively) uncharged;
thus we focus solely on capacitive effects.
A point on the membrane is characterized within the Monge representation by
a height function $h(\rp)$,
where $\rp$ is a two-dimensional in-plane vector in the membrane.

In the electrolyte, the electric potential $\phi$ obeys Poisson's equation
\beq
\label{Poisson_n}
\nabla^2\phi=-\frac{1}{\epsilon}\left(en^+ - en^-\right)=-\frac{2}{\epsilon}\rho\,.
\eeq
Here $e$ is the elementary charge,
$\epsilon$ is the dielectric constant of the electrolyte
and we have introduced {\it half} of the charge density,
\beq\label{rhodef}
\rho=e\frac{\np-\nm}{2}\,.
\eeq
For the sake of simplicity, we assume a symmetric $1:1$ electrolyte, so that
far away from the membrane $n^+=n^-=n^*$, and
the total system is electrically neutral.

The densities of the ion species
obey the Poisson-Nernst-Planck equations
\beq\label{PNP}
\p_t n^\pm+\nabla\cdot\mathbf{j}^{\pm}=0\,,
\eeq
with ionic current densities
\beq\label{currdens}
\mathbf{j}^{\pm}=D\left(-\nabla n^{\pm}\mp n^{\pm}\frac{e}{k_B T}\nabla\phi\right)\,,
\eeq
where $k_B T$ is the thermal energy. For simplicity we consider the case where both ion types
have the same diffusion coefficient $D$ and neglected various corrections for
concentrated solutions~\cite{bazantACIS}.

As boundary conditions (BC), the potential at the electrodes
is externally held at
\beq
\phi\left(z=\pm \frac{L}{2}\right)=\pm \frac{V}{2}\,.
\eeq
This BC is oversimplified for real electrodes,  but captures
the main effects of the electric field, see the discussion in Ref.~\cite{ZBL}.
The distance between the electrodes is assumed to be much larger
than the Debye length, $L\gg\lambda_D=\kappa^{-1}$.
In that case, the bulk electrolyte is quasi-neutral, $n^+=n^-=n^*$, with negligible charge density
(compared to the total salt concentration), so that far from the membrane
\beq
\rho\left(z=\pm \frac{L}{2}\right)
=0\,.
\eeq

The BC at the membrane is crucial to recover the
correct physical behavior.
Let $\mathbf{n}$ be the unit vector normal to the membrane.
We use the Robin-type BC
\beq\label{RobinBC}
&&\dm(\mathbf{n}\cdot\nabla)\phi_{|z=h^+}=\dm(\mathbf{n}\cdot\nabla)\phi_{|z=h^-}=\phi(h^+)-\phi(h^-)\,,
\eeq
where
\beq\label{dmdef}
\dm=\frac{\epsilon}{\epsilon_m}d
\eeq
is a length scale that contains the membrane thickness $d$ and the ratio
of the dielectric constant of the electrolyte, $\epsilon$, and of the membrane, $\epsilon_m$.
This BC was originally developed for electrodes sustaining Faradaic
current \cite{itskovich1977,bazant2005}
or charging capacitively \cite{bazant2004PRE}.
In Refs.~\cite{leonetti:2004,Lacoste_EPJE,ZBL}, this BC was derived, and
was shown to properly account for the jump in the charge distribution which occurs near
the membrane as a result of the dielectric mismatch between the membrane and the surrounding electrolyte.

%%%%%%%%%%%%%%%%%%%%%%%%%%%%%%%%%%%%%%%%%%%%%%%%%%%%%%%%
\section{\label{nlsol} Poisson-Boltzmann approach for a membrane in an external potential}
%%%%%%%%%%%%%%%%%%%%%%%%%%%%%%%%%%%%%%%%%%%%%%%%%%%%%%%%

Here we show how the well-known solution of the Poisson-Boltzmann (PB) equation
for a single charged plate in an electrolyte can be used to describe the
present situation of a capacitive membrane with induced Debye layers in an external potential.

In a steady state situation and when there is no electric current through the membrane,
one obtains from Eqs.~(\ref{PNP}, \ref{currdens})
\beq\label{n_nocurr}
-\nabla n^{\pm}\mp n^{\pm}\frac{e}{k_B T}\nabla\phi=0\,.
\eeq
After a direct integration using the BCs from above, one obtains
\beq
n^{\pm}=n^*\e^{\mp\frac{e}{k_B T}\left(\phi(z)-\frac{V}{2}\right)}\,,
\eeq
and insertion in Poisson's equation
then yields the Poisson-Boltzmann (PB) equation
\beq
\label{PB}
\nabla^2\phi=\frac{2n^*e}{\epsilon}\sinh\left[\frac{e}{k_BT}\left(\phi(z)-\frac{V}{2}\right)\right]\,.
\eeq
Linearization (for $\phi-\frac{V}{2}\ll1$) leads to the well-known Debye-H\"uckel (DH) equation,
\beq
\label{DH}
\nabla^2\phi
=\kappa^2\left(\phi(z)-\frac{V}{2}\right)\,,
\eeq
where
\beq\label{kappa2def}
\kappa^2=\frac{2e^2n^*}{\epsilon k_B T}
\eeq
and $\kappa^{-1}=\lambda_D$ is the Debye length that defines the characteristic length scale for charge relaxation
in the electrolyte.

The nonlinear PB equation (\ref{PB}) for the planar case can be integrated analytically
\cite{andelman},
leading to
\beq
\label{phisol}\phi(z)&=&-\frac{2 k_B T}{e}\ln\left(\frac{1+c\e^{-\kappa z}}{1-c\e^{-\kappa z}}\right)+\frac{V}{2}\,,\\
\label{nsol}n^{\pm}(z)&=&n^*\left(\frac{1\pm c\e^{-\kappa z}}{1\mp c\e^{-\kappa z}}\right)^2\,,
\eeq
for $z>0$.
The expressions for $z<0$
can be obtained using
the symmetry of the system: $\phi(-z)=-\phi(z)$, $n^{\pm}(-z)=-n^{\pm}(z)$.

Just as in the classical PB solution for a single charged plate in contact with an electrolyte,
the non-dimensional parameter $c$ is determined by the BC for the potential at the membrane.
For a flat charged surface
surrounded by an electrolyte \cite{andelman},
$c$ is given by a simple quadratic equation and can be expressed in terms of the ratio of the two characteristic
length scales: the Gouy-Chapman length $b=\frac{2\epsilon k_B T}{e|\sigma|}$ (with $|\sigma|$ the charge
density of the surface) and the Debye length $\lambda_D=\kappa^{-1}$.
In contrast, in the case of a membrane in an electric field, we obtain from Eq.~(\ref{RobinBC})
the following nonlinear equation
\beq\label{ceq}
4\kappa\lambda_m\frac{c}{c^2-1}+\frac{eV}{k_B T}=4\ln\left(\frac{1+c}{1-c}\right)\,.
\eeq
Note that two dimensionless parameters enter this equation: the ratio of electrostatic to thermal energy,
\beq
\bar V=\frac{eV}{k_B T}\,,
\eeq
and the dimensionless parameter~\cite{bazant2005}
\begin{equation}
{\bar\lambda}_m=\kappa\dm=\frac{\dm}{\lambda_D} = \frac{\epsilon\kappa}{\epsilon_m/d}=\frac{C_D}{C_m}\,,
\end{equation}
which quantifies the electrical coupling between the membrane and the Debye layers. More precisely,
depending on whether this parameter is large or small with respect to one,
the capacitance of the diffuse part of the double layer, $C_D=\epsilon\kappa$,
or the capacitance of the membrane, $C_m=\epsilon_m/d$, dominates
the overall voltage drop.

The non-dimensional parameter $c=c(\bar\lambda_m,\bar V)$ which is determined by Eq.~(\ref{ceq})
is related to the potential at the membrane, $\phi(0^+)$,
and
to the charge density at the membrane, $\rho(0^+)$,
by the following relations
\beq
\label{phiofc}
\phi(0^+)=-\frac{4k_B T}{e}{\rm artanh}(c)+\frac{V}{2}\,,\\
\rho(0^+)=en^*\sinh\left[4\,{\rm artanh}(c)\right]\,.
\eeq
From Eq.~(\ref{phiofc}), one can see that the values of $c$ must be restricted
to the interval $[0,1]$. In the limit of small voltages, $\bar V\ll1$, there is a linear relation
between $c$ and the charge distribution at the membrane, since
$c=\frac{\rho(0^+)}{4en^*}=\frac{\epsilon\kappa^2V}{8e n^* (2+\kappa\lambda_m)}$, in
accordance with the calculations based on the DH approximation of Ref.~\cite{ZBL}.

%----------------------------------------------------------------------
\begin{figure}[t]
\begin{center}
\includegraphics[width=13cm]{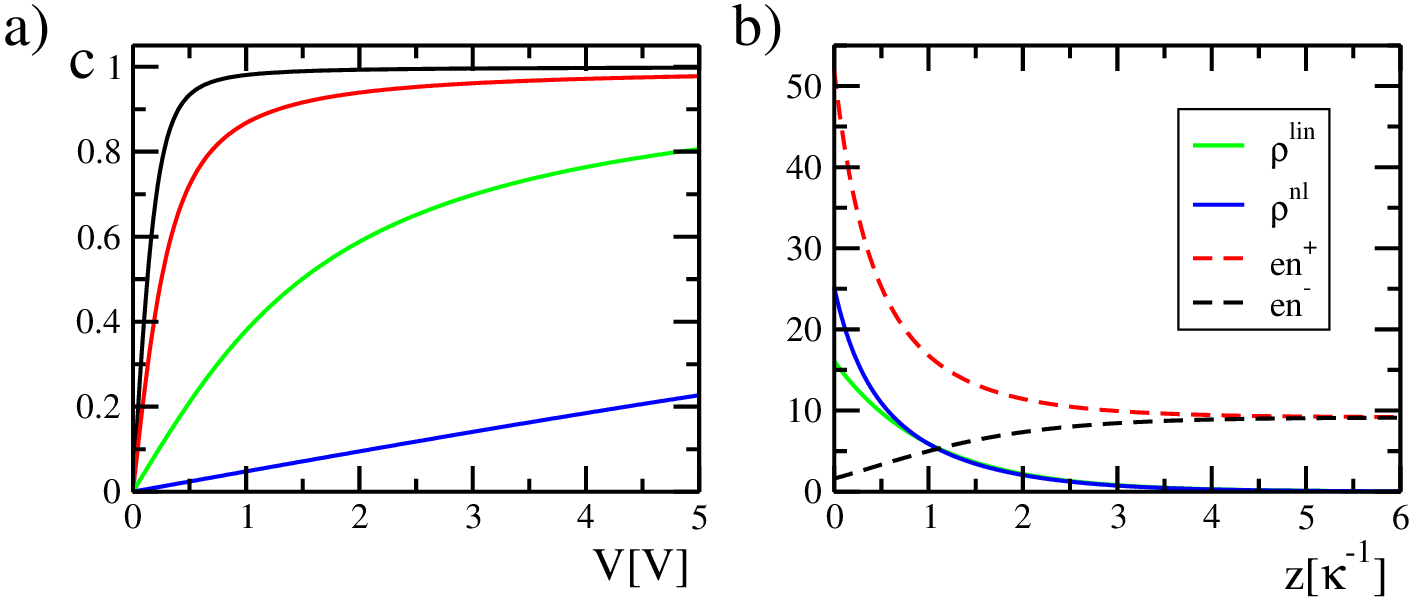}
\caption{\label{fig1} a) Solution of Eq.~(\ref{ceq}) for different amounts of salt:
$\kappa=10^{6}{\rm m}^{-1}$ (pure water; black),
$\kappa=10^{7}{\rm m}^{-1}$ (red), $\kappa=10^{8}{\rm m}^{-1}$ (green),
$\kappa=10^{9}{\rm m}^{-1}$ (blue).
b) Charge distribution (in units
of ${\rm C m}^{-3}$) for $z>0$ : comparison of the nonlinear solution (blue) with the
linearized Debye-H\"uckel solution (green). The asymmetric distribution
of positive (red dashed) and
negative charges (black dashed) for the nonlinear solution is also shown.
Parameters used:
$\epsilon=80\epsilon_0$ (water) and $\epsilon_m=2\epsilon_0$ for the membrane.
As the membrane thickness is typically $d=5{\rm nm}$ this leads to
$\dm=\frac{\epsilon}{\epsilon_m}d =200{\rm nm}$.
For b) we used $\kappa=10^{6}{\rm m}^{-1}$, corresponding to $en^* \simeq 9.16\,{\rm Cm}^{-3}$, and
$V=0.1V$.
}
\end{center}
\end{figure}
%--------

Fig.~\ref{fig1}a) shows the solution $c$ of Eq.~(\ref{ceq}) as a function of external voltage
for various values of $\kappa$, which corresponds to varying the amount of salt since $\kappa\propto\sqrt{n^*}$.
Clearly, for low salt, the linear approximation remains valid only for rather small
voltages (for instance it holds only for $V\lesssim 0.1{\rm V}$ in pure water),
while for high salt $V=5{\rm V}$ is still in the linear regime.
Fig.~\ref{fig1}b) shows a comparison of the charge distribution for the nonlinear PB (blue)
and the linear DH solution (green). As expected, the figure shows that the DH approximation
underestimates the surface charge on the membrane layers as compared to the PB calculation.
The figure also shows the distribution of the positive and negative ions, which both
tend to $n^*$ far from the interface as a result of electroneutrality.

Although the present situation differs from the case of a single charged plate in an PB electrolyte,
the structure of the solutions Eqs.~(\ref{phisol}, \ref{nsol})
is very similar in both problems. Because of this, there is an equivalent to the
{\it Contact theorem} \cite{andelman}, which relates in the single charged plate problem
the surface charge density to the limiting value of the potential/ionic density at the plate:
namely, one can give the effective surface charge $\sigma_{\rm eff}$
for the charged plate problem that creates the same voltage/charge distributions
as the capacitive membrane in the external field.
This effective surface charge reads
\beq
\sigma_{\rm eff}^2=4\epsilon n^* k_B T\left[\cosh\left(\frac{e}{k_B T}(\phi(0^+)-V/2)\right)-1\right]\,.
\eeq

%%%%%%%%%%%%%%%%%%%%%%%%%%%%%%%%%%%%%%%%%%%%%%%%%%%%%%%%
\section{\label{corr_surf}Corrections to membrane elastic moduli}
%%%%%%%%%%%%%%%%%%%%%%%%%%%%%%%%%%%%%%%%%%%%%%%%%%%%%%%%

{\bf Surface tension.} The electrostatic corrections
to the surface tension can be calculated directly from the stresses
acting on the membrane in the flat configuration (also called the base state)
as explained in Refs.~\cite{widom,ZBL}.
The total stress tensor reads
\beq\label{stresstensor}
\tau_{ij}=-p\delta_{ij}+\eta\left(\p_i v_j+\p_j v_i\right)
+\epsilon\left(\hspace{-1mm}E_i E_j -\frac{1}{2}\delta_{ij}E^2\hspace{-1mm}\right)\,
\eeq
which contains the pressure,
the viscous stresses in the surrounding fluid (the electrolyte)
and the Maxwell stress due to the electrostatic field. $\eta$ is the viscosity
of the electrolyte and $\mathbf{v}$ its velocity field. The electric field is given by
$\mathbf{E}=-\nabla\phi$.

In the base state, the electric field is oriented in $z$-direction,
and the condition $\nabla\cdot\tau=0$ implies that
$\p_z p=\frac{\epsilon}{2}\p_z\left(\left(\p_z\phi\right)^2\right)=-2\rho\p_z\phi$.
Using Eqs.~(\ref{phisol}, \ref{nsol})
and imposing $p(z\rightarrow\infty)=0$, this is readily solved by
\beq
\label{pressure_base}
p(z>0)=16n^* k_B T c^2\frac{\e^{2\kappa z}}{\left(c+\e^{\kappa z}\right)^2\left(c-\e^{\kappa z}\right)^2}\,,
\eeq
and similarly with $z\rightarrow -z$ for $z<0$.

Let us call $S$ a closed surface englobing the membrane
with the normal vector field ${\bf n}$.
The force acting
on the surface $S$ in the ${\bf x}$-direction (chosen to be the direction of the lateral stress)
can be calculated
from the stress tensor as
$F_x=\int_S {\bf x} \cdot \tau \cdot {\bf n} \,\,dS$.
Since $\tau$ is divergence free, the surface $S$ can be
deformed,
for convenience to a cube of size $L$, and it is easy to see that the integral
is non-zero only on the faces of the cube with the
normal along $\pm{\bf x}$. With $dS=L dz$ and $\Delta
\Sigma=F_x/L$ for
${\bf n}=+{\bf x}$, we arrive at
\beq \label{new delta sigma}
\Delta \Sigma=\int_{-L/2}^{L/2} \tau_{xx}(z) dz\,.
\eeq
From Eq.~(\ref{stresstensor}), one has
$\tau_{xx}(z)=-p(z)-\frac{\epsilon}{2} (\partial_z \phi)^2$,
where $\phi(z)$ and $p(z)$ are the potential and the pressure in the base state given above.
Upon integration (using $L\kappa\gg1$),
one obtains the corrections to the surface tension, as a sum of two terms.

First, there is the \emph{external contribution} due to the Debye layers,
\beq\label{dsk}
\Delta\Sigma_\kappa=-\frac{32 n^* k_B T}{\kappa}\frac{c^2}{1-c^2}\,.
\eeq
Second, there is the \emph{internal contribution} due to the electric field
inside the membrane (cf. Ref.~\cite{ZBL}),
which is given by $E_0^m=-\frac{1}{d}\left(\phi(0^+)-\phi(0^-)\right)$.
That correction to the surface tension is
$\Delta \Sigma_m=-\epsilon_m d \left(E_0^m\right)^2$, or explicitly
\beq\label{dsm}
\Delta \Sigma_m=-\frac{\epsilon_m}{d}\left[-\frac{4k_B T}{e}\ln\left(\frac{1+c}{1-c}\right)+V\right]^2\,.
\eeq
Note that both corrections to the surface tension are {\it negative}, which means that these corrections
can lead to an instability as soon as the total surface tension
$\Sigma_{tot}=\Sigma_0+\Delta\Sigma_\kappa+\Delta \Sigma_m$
(which is the sum of the
bare tension $\Sigma_0$ plus the above corrections) becomes negative \cite{pierre}.

{\bf Bending modulus.} To obtain the correction to the bending modulus
we perform, as detailed in Ref.~\cite{ZBL} for the DH case,
a calculation
of the potential at first order in the membrane height.
Then, by solving the hydrodynamics problem
of the electrolyte around the membrane (in Stokes approximation),
one determines the pressure and obtains the total stress tensor.
The growth rate of membrane fluctuations, $s(\kpm)$, where $\kp$ is the wave vector in the membrane plane,
is then determined by imposing that
the discontinuity of the normal-normal component of the total stress tensor at the
membrane has to equal the membrane restoring force:
\beq
\label{BC_stress_normal}
-\left[(\tau_{zz,1}(z=0+)-\tau_{zz,1}(z=0-)\right]
= -\frac{\partial F_H}{\partial h(\rp)}\,.
\eeq
Here $F_H$ is the standard Helfrich free energy,
\beq \label{mb free energy1}
F_H=\frac{1}{2} \int d^2 \rp [ \Sigma_0 \left( \nabla h \right)^2 + K_0 \left( \nabla^2 h
\right)^2  ],
\eeq
and $\Sigma_0$ and $K_0$ are the bare surface tension and the bare bending modulus of the membrane,
respectively. Expanding the left hand side of Eq.~(\ref{BC_stress_normal}) in powers of $\kp$
yields the growth rate of the form
\beq\label{disp_main}
\hspace{-5mm}\eta \kpm s(\kpm)=
-\frac{1}{4}\left(\Sigma_0+\Delta\Sigma_\kappa+\Delta\Sigma_m\right)\kpm^2
-\frac{1}{4}\left(K_0+\Delta K_\kappa+\Delta K_m\right)\kpm^4\,.\quad\quad\quad
\eeq

Details of the calculations can be found in \ref{app1}.
The surface tension corrections calculated above in
Eqs.~(\ref{dsk}, \ref{dsm}) can be recovered by this method,
which provides a self-consistency check.
For the bending modulus one obtains
\beq
\label{Kk}\Delta K_\kappa&=&\frac{8 n^* k_B T}{\kappa^3}\frac{c^2(3-c^2)}{1+c^2}\,,\\
\Delta K_m&=&-\epsilon_m\left(E^0_m\right)^2\left[-\frac{d^3}{12}
+\frac{2k_B T}{E^0_m e\kappa}\frac{c(1-c^2)}{1+c^2}d\right]\,,
\eeq
for the \emph{external contribution} due to the Debye layers and
the \emph{internal contribution} due to the voltage drop at the membrane, respectively.
The field inside the membrane is given by
$E_0^m=-\frac{1}{d}\left[-\frac{4k_B T}{e}\ln\left(\frac{1+c}{1-c}\right)+V\right]$.

%%%%%%%%%%%%%%%%%%%%%%%%%%%%%%%%%%%%%%%%%%%%%%%%%%%%%%%%
\section{\label{discuss}Discussion}
%%%%%%%%%%%%%%%%%%%%%%%%%%%%%%%%%%%%%%%%%%%%%%%%%%%%%%%%

Let us now discuss the nonlinear electrostatic effects on the membrane elastic moduli
in the limits of low and high applied voltages.

%--------------------------------------------------------
{\bf Low voltage regime.}
In the low voltage limit, $\bar V\ll 1$, a solution of
Eq.~(\ref{ceq}) to linear order in $c$, yields
\beq
c=\frac{\bar V}{4(2+\bar\lambda_m)}=\frac{1}{4(2+\lambda_m\kappa)}\frac{eV}{k_BT}=\frac{\rho(0^+)}{4en^*}\,.
\eeq
Here $\rho(0^+)$ is half the charge density at the membrane, corresponding to
the quantity called $\rho_m$ in Ref.~\cite{ZBL} for the DH case.
Expanding $\phi(z)$ and $\rho(z)$, as well as the corrections to the moduli
$\Delta\Sigma_\kappa$, $\Delta \Sigma_m$, $\Delta K_\kappa$ and $\Delta K_m$
for small $c$,
one exactly recovers all of the results given in Ref.~\cite{ZBL}.
Specifically, all corrections to the moduli scale quadratically with the external voltage,
$\propto V^2$.
%The applied electric field modifies
%the electrostatic energy of the electrolyte and of the membrane in a way which is
%a quadratic with respect to the applied voltage, since
This is due to the fact that both the potential and the induced charge
are proportional to the applied voltage.

%---------------------------------------------------------------------------
{\bf High voltage regime.} In the opposite limit, $\bar V\gg1$ implies $c\rightarrow 1$.
Introducing $\alpha=1-c$ and expanding Eq.~(\ref{ceq}) for small $\alpha$,
one gets
$-\frac{2\bar{\lambda}_m}{\alpha}+4\ln\alpha+\bar{V}+\bar{\lambda}_m-4\ln2=0$.
For high values of $\bar{V}$ one can neglect the last two (constant) terms and gets
\beq
c
=1-\frac{\frac{1}{2}\bar{\lambda}_m}{W\left(\frac{1}{2}\bar{\lambda}_m\e^{\bar{V}/4}\right)}
\simeq1-\frac{\frac{1}{2}\bar{\lambda}_m}{\bar{V}/4}\,,
\eeq
where $W(x)$ is Lambert's function, i.e. $y=W(x)$ is the solution of $ye^y=x$.
%
%----------------------------------------------------------------------
\begin{figure}[t]
\begin{center}
\includegraphics[width=7.5cm]{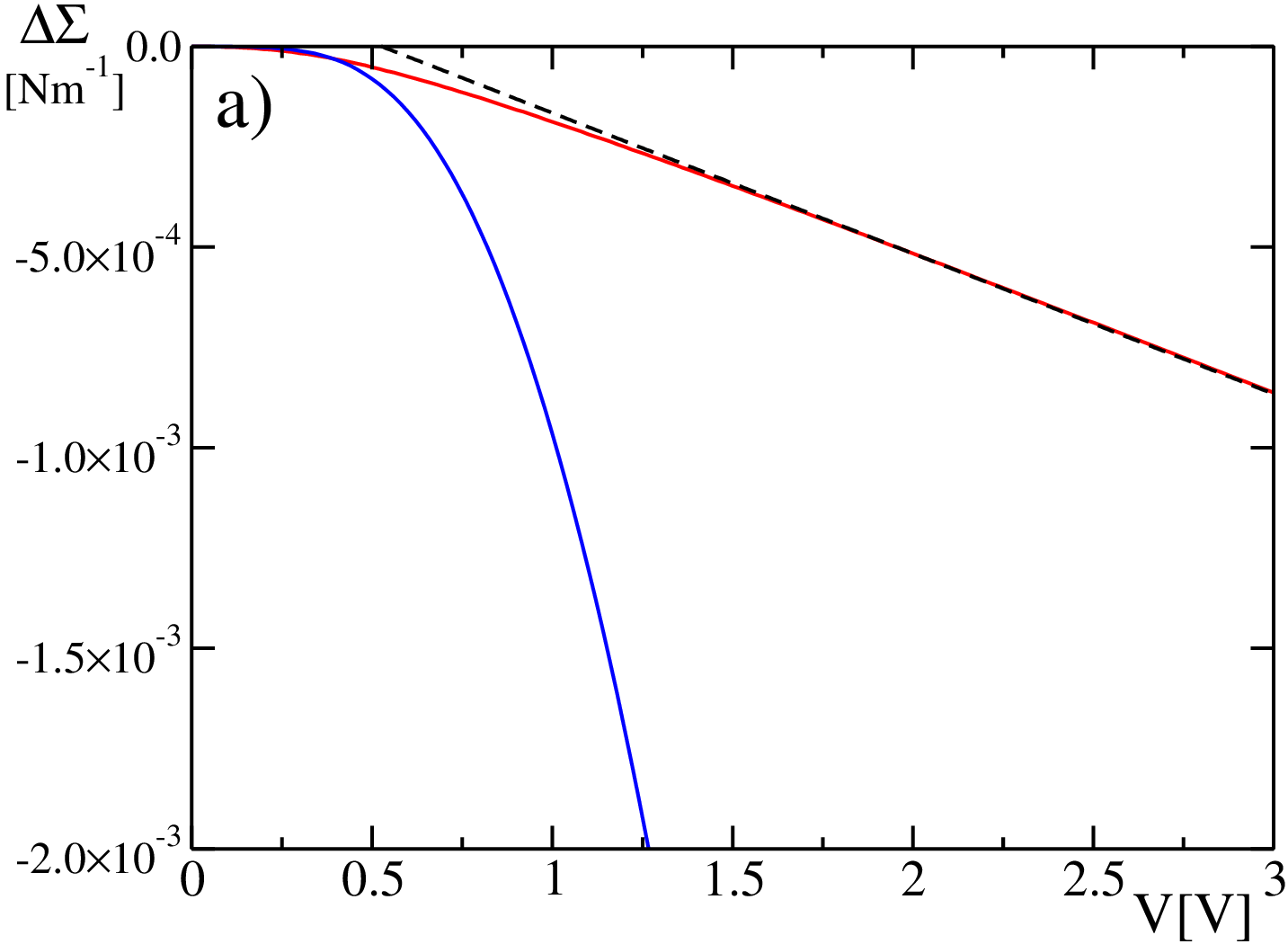}
\includegraphics[width=7.5cm]{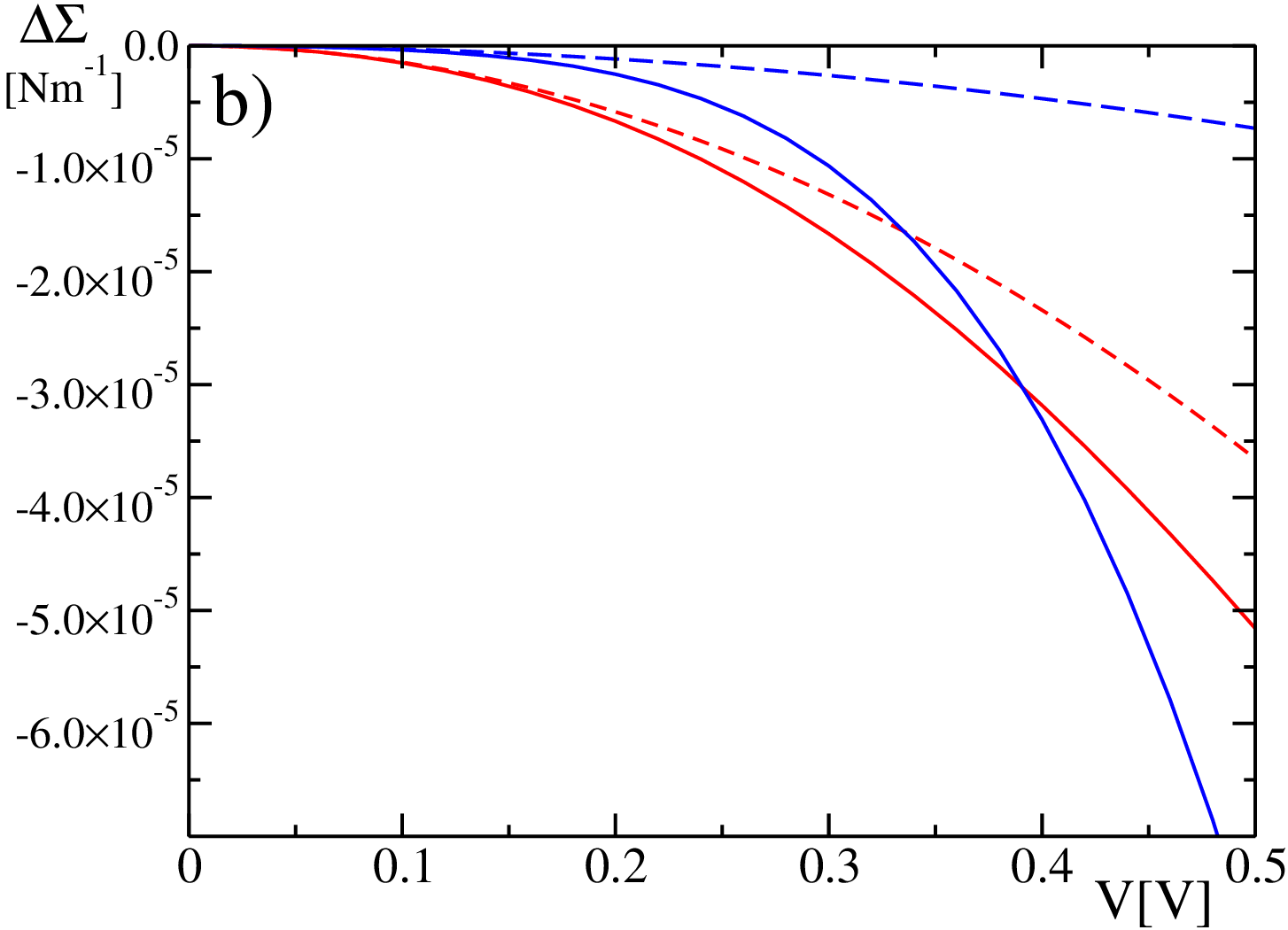}
\caption{\label{fig2} Corrections to the surface tension for $\kappa=10^6{\rm m}^{-1}$ (pure water)
as a function of voltage.
Corrections from the Debye layer are in red, corrections due to the voltage drop at the
membrane are in blue.
a) Nonlinear PB result. One sees that the correction due to
the Debye layer scales linearly in the voltage for high voltages (cf. the black dashed straight line),
while the contribution from inside the membrane scales quadratically in the voltage.
b) Comparison of the nonlinear PB result (solid lines) with the linear DH result (dashed lines).
}
\end{center}
\end{figure}
%--------
%
%
As a result, the nonlinear electrostatics strongly effects the corrections to the moduli
from the Debye layers. In fact, the external contribution to the surface tension scales as
\beq
\Delta\Sigma_\kappa(\bar{V}\gg1)\propto-\frac{c^2}{1-c^2}
%\propto-\frac{\left(\bar{V}-2\bar{\lambda}_m\right)^2}{4\bar{\lambda}_m\left(\bar{V}-\bar{\lambda}_m\right)}
\rightarrow-\frac{\bar{V}}{4\bar{\lambda}_m}\,
\eeq
instead of
$\Delta\Sigma_\kappa(\bar{V}\ll1)\propto-\frac{\bar{V}^2}{(2+\bar{\lambda}_m)^2}$ :
in the high voltage regime, the external surface tension correction
scales linearly with the voltage
instead of quadratically.
The crossover from $\bar V^2$ to $\bar V$ is salt dependent,
i.e. depends on the value of $\bar{\lambda}_m=\lambda_m \kappa$, cf.~Fig.~\ref{fig1}.
In contrast, the external surface tension correction
$\Delta\Sigma_m$ remains a quadratic function of the
applied voltage.

Fig.~\ref{fig2} displays both corrections to the surface tension
as a function of voltage for low salt (pure water). Fig.~\ref{fig2}a) shows that
due to the crossover to a linear voltage dependence for the external surface tension correction,
rapidly $\Delta\Sigma_m$ (the blue curve) wins and $\Delta\Sigma_\kappa$ (red curve) becomes
negligible for $V\gtrsim 1.5V$.
Fig.~\ref{fig2}b) shows a comparison of the nonlinear PB result (solid lines)
with the linear DH result (dashed lines).
For $V\lesssim 0.2V$, there is agreement and the external contribution dominates.
However, for higher voltages the linearized DH
solution becomes completely misleading: it predicts that $|\Delta\Sigma_\kappa|>|\Delta\Sigma_m|$ for all voltages
(and both proportional to $V^2$), while already above $V\gtrsim0.4V$ the internal contribution
exceeds the external contribution.

\begin{figure}[t]
\begin{center}
\includegraphics[width=7.5cm]{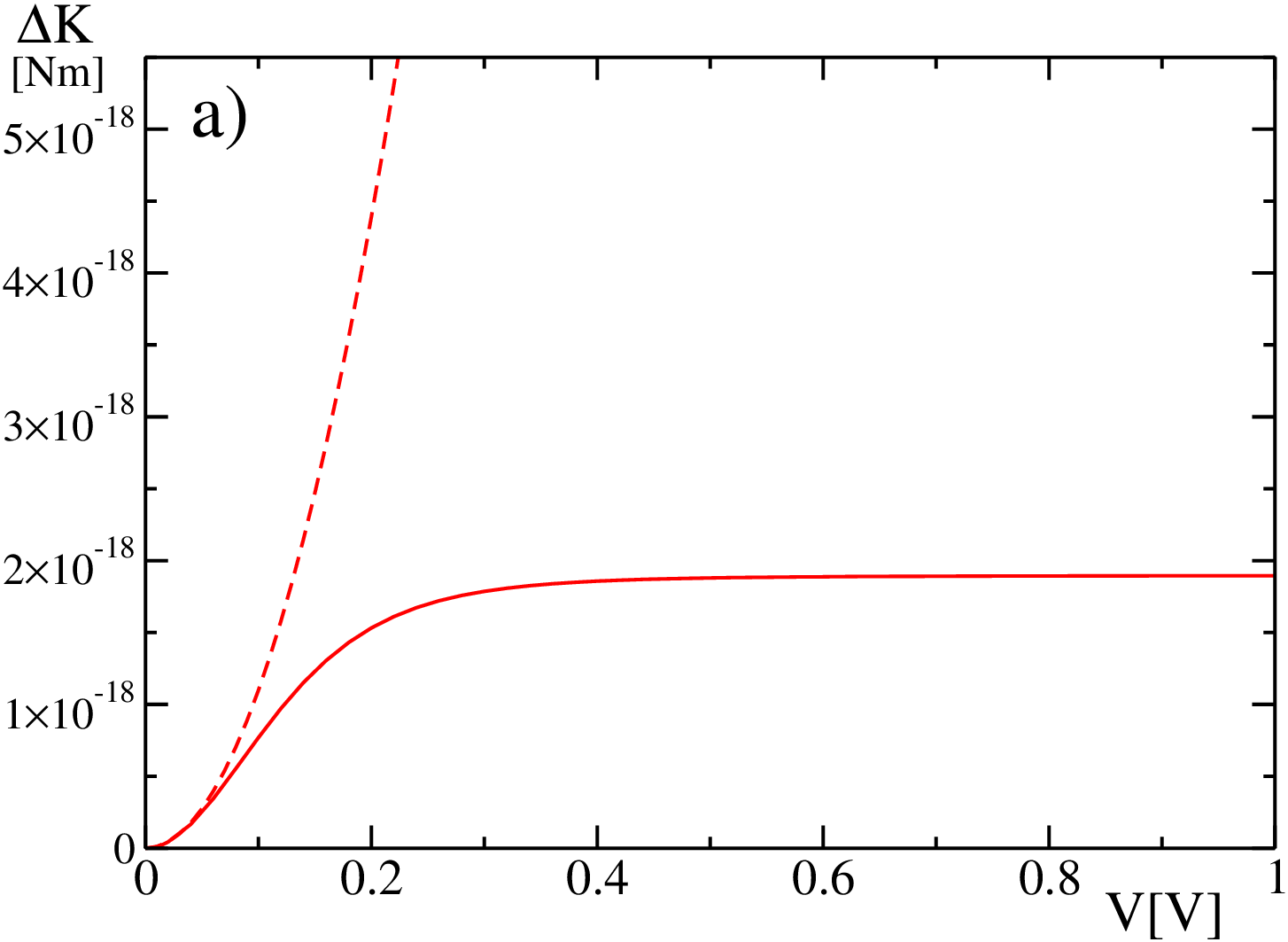}
\includegraphics[width=7.5cm]{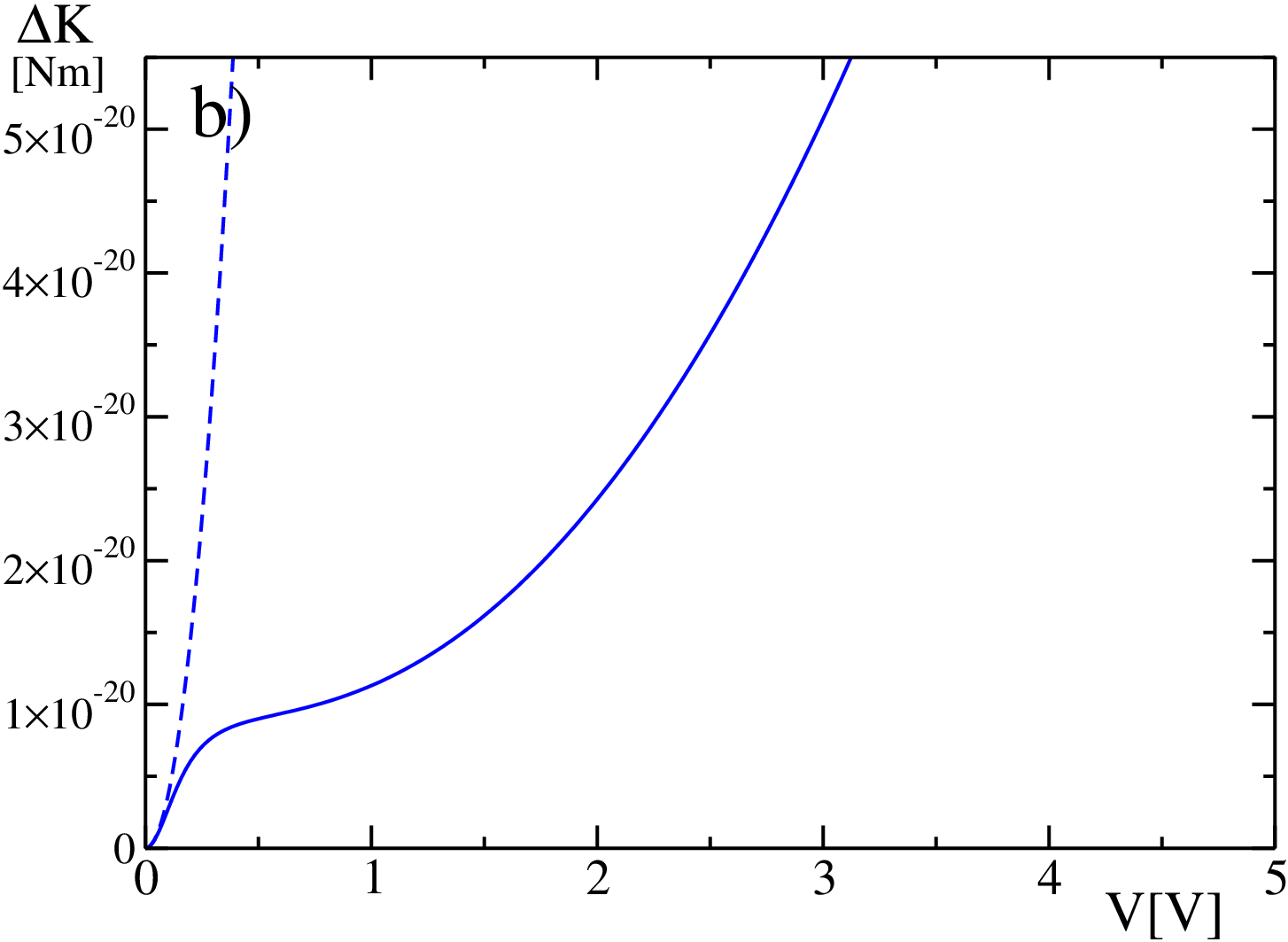}
\caption{\label{fig3} Corrections to the bending modulus for $\kappa=10^6{\rm m}^{-1}$ (pure water)
as a function of voltage.
a) Corrections from the Debye layer.
b) Corrections due to voltage drop at the membrane.
Nonlinear PB calculations are solid, linearized DH calculations are dashed
(note the different scales on the left and right panels).
}
\end{center}
\end{figure}

For the bending modulus corrections, the differences between the DH and PB models are even more striking:
As a function of the applied voltage, the external contribution levels off at a constant value
\beq
\Delta K_\kappa(\bar{V}\gg1)=\frac{8 n^* k_B T}{\kappa^3}\frac{c^2(3-c^2)}{1+c^2}\rightarrow\frac{8 n^* k_B T}{\kappa^3}\,.
\eeq
In contrast, the internal contribution $\Delta K_m$ continues to grow
quadratically in the high voltage limit.
Fig.~\ref{fig3}a) and b) show comparisons of the nonlinear PB solutions (solid)
and the linear DH solutions (dashed) for both contributions to the bending modulus.
Note, however, the different scales:
Although the external contribution saturates, see Fig.~\ref{fig3}a),
this value is larger by two orders of magnitude than the still growing value
from the internal contribution, cf.~Fig.~\ref{fig3}b).
In conclusion, for pure water and voltages $\simeq0.2-5V$, the total correction to
the bending modulus will appear constant within this voltage interval. Only for still
higher voltages the quadratic growth due to the internal contribution $\Delta K_m$ will dominate.

%----------------------------------------------------------------------

\begin{figure}[t]
\begin{center}
\includegraphics[width=7.5cm]{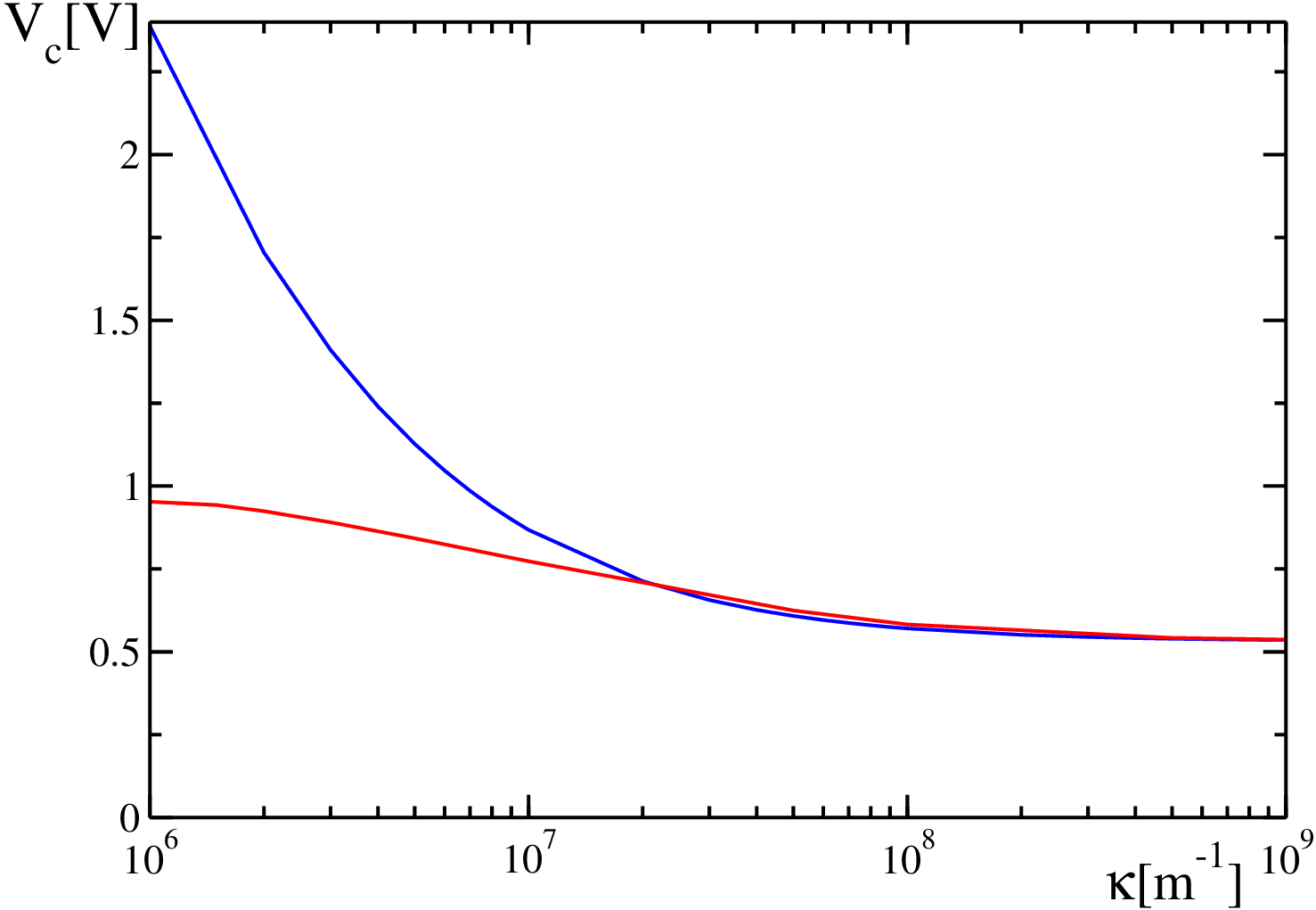}
\caption{\label{fig4} Threshold voltage $V_c$ for the membrane undulation instability as
a function of salt (i.e. $\kappa$). Nonlinear PB result (red curve) vs. linear DH result (blue curve).
Parameters: $\kappa=10^6{\rm m}^{-1}$ (pure water);
bare surface tension $\Sigma_0=10^{-3} {\rm N m}^{-1}$; $\lambda_m=200 {\rm nm}$.
}
\end{center}
\end{figure}

{\bf Membrane instability.} Let us briefly discuss the consequences for the membrane undulation instability.
As mentioned above, an instability develops in this system when the total surface tension,
the sum of the bare value and the electrostatic corrections, becomes negative.
The threshold value for the voltage, $V_c$, has been calculated in Ref.~\cite{ZBL} within
linearized DH theory and reads (in case of a non-conductive membrane)
\beq\label{Vc}
V_c^2=\frac{\Sigma_0 d\left(2+\kappa\lambda_m\right)}{\kappa\left(\kappa\epsilon_m\lambda_m^2+\epsilon d\right)}\,.
\eeq
This curve is shown in Fig.~\ref{fig4} as the blue line.
The red line, in contrast, shows the nonlinear PB result given by numerical solution
of $0=\Sigma_{tot}=\Sigma_0+\Delta\Sigma_\kappa+\Delta \Sigma_m$ with Eqs.~(\ref{dsk}, \ref{dsm})
for the surface tension corrections and $c(\bar\lambda_m,\bar V)$ calculated from Eq.~(\ref{ceq}).
In the high salt limit, the Debye layers shrink to zero and as a result the external contribution vanishes. For the internal contribution, the inside field is exactly calculated from the DH approach,
as can be seen from the behavior of the parameter $c$ in Figure \ref{fig1}a.
Thus, both curves merge and attain the limiting value given by Eq.~(\ref{Vc}),
$V_c(\kappa\rightarrow\infty)=\sqrt{\frac{\Sigma_0 d}{\epsilon_m}}$.
In contrast, for low salt, the linear result overestimates the threshold value, since it
underestimates the induced charges at the membrane, as shown in Fig.~\ref{fig1}b).
%
%----------------------------------------------------------------------

In addition to the threshold voltage for the instability, the most unstable
wave number associated to the instability will be affected by the electrostatic nonlinearities of the PB equation.
This wave vector corresponds to the maximum of the growth rate given by Eq.~(\ref{disp_main})
with respect to $\kpm$. In the low salt regime, since the system is more unstable in the presence
of electrostatic nonlinearities, cf. Fig.~\ref{fig4}, in general the wave vector
will be increased as compared to the predictions from the DH approach.

%================================
\section{\label{conc}Conclusions}
%================================

We have investigated the nonlinear electrostatic effects
of an external DC electric field on a {\it purely capacitive} membrane,
which is non-conductive for the ions and bears no fixed charges, in an electrolyte.
We have calculated in the nonlinear Poisson-Boltzmann regime
the corrections to the membrane elastic moduli --
both the external ones due to the Debye layers surrounding the membrane
and the internal corrections due to the electric field inside the membrane.
Strong deviations from the linear Debye-H\"uckel behavior have been found
in the low salt regime at already rather moderate voltages.
In particular we have shown that the
external contribution to the surface tension crosses over from a
quadratic dependence on the externally applied voltage as predicted by the linearized theory
to a linear voltage dependence.
In contrast, the internal contribution remains quadratic and becomes dominating at high voltages.
The external contribution to the
bending modulus even saturates for high voltages, while
the internal contribution remains quadratic in voltage.

In addition, our work confirms that the surface tension still grows in absolute value with
the voltage, which means that the membrane undulation instability present in the DH theory
(due to an effectively negative surface tension) persists in the nonlinear PB regime.
The nonlinearities affect the threshold in voltage and the characteristic
wavelength of the instability.

The method presented here can serve as a starting point for further extensions.
One example would be to include fixed charges in addition to
induced charges, similarly as studied in \cite{sachs}.
Another possible direction would be to improve the description of the membrane,
cf.~the recent work in Ref.~\cite{olmsted} where thickness fluctuations of the membrane
and fluctuations of the lipid dipole orientations within the membrane are accounted
for in a comprehensive continuum model for a membrane in a normal DC electric field.
Other possible extensions could include other types of non-linear effects,
for example due to membrane elasticity, due to inclusions of proteins such as ion channels or pumps
in the membrane \cite{faris}, and also various relevant non-equilibrium effects,
coupling electrostatics and hydrodynamics as in induced charge electro-osmosis \cite{bazant:2004}.
Finally, in the biological context, the heterogeneity of the bilayer composition
is another feature which is beyond the present model and which is likely to be important.

%================================
\ack
%================================
We would like to acknowledge stimulating discussions with L. Malaquin,
T. Charitat and J-F. Joanny.
F.Z. acknowledges financial support from the German Science
Foundation (DFG). F.Z. is on leave from the University of Bayreuth, Germany.
%================================

\begin{appendix}
\section{Details of calculation of bending modulus}
\label{app1}

As this is the extension to nonlinear electrostatics of earlier work \cite{ZBL},
we only sketch the calculations. To solve the electrostatics problem to first order
in the membrane height, one linearizes in $h$ by writing
\beq
\phi(\kp,z)=\phi_0(z)+\phi_1(\kp,z)\,,\nonumber\\
\rho(\kp,z)=\rho_0(z)+\rho_1(\kp,z)\,,\nonumber\\
n^\pm(\kp,z)=n_0^\pm(z)+n^\pm_1(\kp,z)\,,\nonumber
\eeq
where $\phi_0$, $n_0^\pm$ are the base state solutions (flat membrane) as given in the main part
and $\rho_0=\frac{e}{2}(n_0^+-n_0^-)$. Quantities with subscript 1 are of order $h$. We used
the definition of the Fourier transform for the in plane vector $\rp$,
$f(\kp,z)=\int d\rp e^{-i\kp\cdot\rp}f(\rp,z)$.

Assuming zero current through the membrane, the Poisson-Nernst-Planck equation
linearized in $h$ has solutions
$n_1^+=-n_0^+\phi_1$ and $n_1^-=n_0^-\phi_1$.
Insertion into the PB equation yields to linear order in $h$
\beq
\left(\p_z^2-\kpm^2\right)\phi_1
=\kappa^2\,\frac{\left(1+6c^2e^{-2\kappa z}+c^4e^{-4\kappa z}\right)}{\left(1-c^2e^{-2\kappa z}\right)^2}\,\phi_1\,,
\eeq
which is solved (with BC $\phi_1(z\rightarrow\infty)=0$) by
\beq\label{phi1}
\phi_1=A\,\sqrt{\left(1-4\frac{\kappa^2}{\kpm^2}\frac{c^2e^{2\kappa z}}{\left(e^{2\kappa z}-c^2\right)^2}\right)
\frac{e^{2\kappa z}-\frac{(l-\kappa)^2}{\kpm^2}c^2}{e^{2\kappa z}-\frac{(l+\kappa)^2}{\kpm^2}c^2}}\,\cdot\,e^{-lz}\,,
\eeq
where we have introduced $l^2=\kpm^2+\kappa^2$ (note that for $c\ll1$ one regains
a simple exponential, $\phi_1=A'e^{-lz}$, as in Ref.~\cite{ZBL}).
The constant $A$ can be obtained from the BC, Eq.~(\ref{RobinBC}), at order $h$
\beq\label{BCOh}
\lambda_m\left[\left(\p_z^2\phi_0\right)_{|z=0} \cdot h +\left(\p_z\phi_1\right)_{|z=0}\right]=\phi_1(0^+)-\phi_1(0^-)\,.
\eeq
As by symmetry $\phi(z)=-\phi(-z)$ and $\phi_1\propto h$, one
has $\phi_1(z)=\phi_1(-z)$ and Eq.~(\ref{BCOh}) simplifies to
$\left(\p_z\phi_1\right)_{|z=0}=-\left(\p_z^2\phi_0\right)_{|z=0} \cdot h$.
$A$ is thus independent of $\lambda_m$.
The full expression for $A$ is not needed, since later on we expand in $\kpm$.

Since we study the case without ionic current through the membrane,
the hydrodynamics problem (cf.~Ref.~\cite{ZBL}) around the membrane  is trivial
and one gets
\beq
v_z=h(\kp)s(\kpm)\left(1+\kpm z\right)e^{-\kpm z}
\eeq
for the normal component of the velocity,
as induced by a pure membrane bending mode \cite{BrochardLennon}.
Here $s(\kpm)$ is the growth rate of the membrane fluctuations.
The pressure is given by (in incompressible Stokes approximation)
\beq\label{press_dim}
p=-\eta\p_z v_z+\frac{\kp\cdot\fp}{i\kpm^2}+\frac{\eta}{\kpm^2}\p_z^3v_z\,.
\eeq
Herein enters the bulk force due to the electric field acting on the charge distribution,
which reads
$\mathbf{f}=-2\rho\nabla\phi
=-2\rho_0\nabla\phi_1-2\rho_1\nabla\phi_0+\mathcal{O}(h^2)$
with perpendicular component
\beq
\label{fp}\fp&=&-2\rho_0(z)i\kp\phi_1(\kp,z)\,.
\eeq

The total normal stress at the membrane to linear order in $h$ is, cf.~Eq.~(\ref{BC_stress_normal}),
\beq\label{full_stress}
\tau_{zz,1}=\hspace{-1mm}
\left[-p+2\eta\p_z v_z+\frac{\epsilon}{2}\left(\p_z \phi\right)^2
-\frac{\epsilon_m}{2}\left(\p_z \phi^m\right)^2\right]_{|z=h}\,.
\eeq
We now have to calculate the normal stress discontinuity at the membrane
up to fourth order in $\kpm$
and to balance it with the membrane restoring force.
With the abbreviation $[f]_0=f(0^+)-f(0^-)$ the stress discontinuity can be written as
\beq\label{stress_disc}
\hspace{-2.5cm}[\tau_{zz,1}]_{z=0}\hspace{-1mm}&=&\hspace{-1mm}-[p]_{z=0} + 2\eta[\p_z v_z]_{z=0}
+\epsilon\left[(\p_z\phi_0)(\p_z\phi_1)\right]_{z=0}
-\epsilon_m\left[(\p_z\phi^m_{0})(\p_z\phi^m_{1})\right]_{z=\pm d/2}\,\hspace{-1mm}.\quad\quad\quad
\eeq
The balance reads
\beq\label{balancenew}
-[\tau_{zz,1}]_{z=0}
= -\frac{\partial F_H}{\partial h(\rp)}
=\left(-\Sigma_0\kpm^2-K_0\kpm^4\right)h(\kp)\,
\eeq
and yields the growth rate of membrane fluctuations $s(\kpm)$
of the form
\beq\label{disp_fin}
\eta \kpm s(\kpm)=
-\frac{1}{4}\left(\Sigma_0+\Delta\Sigma\right)\kpm^2
-\frac{1}{4}\left(K_0+\Delta K\right)\kpm^4\,.
\eeq

A little care has to be taken for the correction due to the internal field, cf. Ref.~\cite{ZBL}.
Note that we accounted for the finite membrane thickness $d$ in Eq.~(\ref{stress_disc}) above.
As the internal field is constant, and due to symmetry,
one can write
\beq\label{inside}
\epsilon_m\left[(\p_z\phi_{0}^{m})(\p_z\phi_{1}^{m})\right]_{|z=+d/2}
=-2\epsm E_{0,m}\left(\p_z\phi_{1}^{m}\right)_{|z=+d/2}.
\eeq
The potential inside the membrane can be written (using again the symmetry, as well as $\rho=0$ inside; cf.
also Ref.~\cite{Lacoste_EPJE} for details)
\beq
\phi_{1}^{m}(\kpm,z)=\phi_{1}^{m}\left(\kpm,\frac{d}{2}\right)
\frac{e^{\kpm d/2}}{e^{\kpm d}+1}\left(e^{\kpm z}+e^{-\kpm z}\right)\,.
\eeq
$\phi_{1}^{m}\left(\kpm,\frac{d}{2}\right)$ can be calculated approximately
by using the outside solution, Eq.~(\ref{phi1}), and imposing the BC at the membrane,
leading to
\beq
\phi_{1}^{m}\left(\kpm,\frac{d}{2}\right)=
\phi_{1}\left(\kpm,\frac{d}{2}\right)
-h(\kpm)\left(\p_z\phi_{0}^{m}-\p_z\phi_{0}\right)_{|z=d/2}\,.
\eeq
On the right hand side, now all quantities are known.
Finally, one expands Eq.~(\ref{inside}) and all the other quantities entering the
total stress discontinuity, Eq.~(\ref{full_stress}), up to fourth order in $\kpm$.
From Eq.~(\ref{balancenew}) one then obtains the growth rate $s(\kpm)$ of membrane fluctuations.

\end{appendix}

\end{document}